\documentclass[12pt,preprint]{aastex}
\newcommand{\nraoblurb}{The National Radio Astronomy Observatory is
a facility of the National Science Foundation operated under cooperative
agreement by Associated Universities, Inc.}

\newcommand{\percc}{$\,{\rm cm^{-3}}$}

\newcommand{\kms}{\ensuremath{\,{\rm km\,s}^{-1}}}

\newcommand{\degper}{\ensuremath{\rlap.{^{\circ}}}}

\newcommand{\hii}{{\rm H\,}{{\sc ii}}}       
\newcommand{\he}[1]{$^{#1}{\rm He}$}

\newcommand{\ci}{{\rm C\,}{{\sc i}}}
\newcommand{\cii}{{\rm C\,}{{\sc ii}}}
\newcommand{\hethree}{\ensuremath{{}^3{\rm He}}}
\newcommand{\hefour}{\ensuremath{{}^4{\rm He}}}
\newcommand{\hepthree}{\ensuremath{{}^3{\rm He}^+}}

\newcommand{\halpha}{\ensuremath{\langle{}\,{\rm H}{}\,\rangle{}\,\alpha}}
\newcommand{\nexpo}[2]{\ensuremath{{#1}\times 10^{#2}}}

\shorttitle{Galactic Radio Recombination Lines}
\shortauthors{Quireza et al.}

\begin{document}

\title{Radio Recombination Lines in Galactic H\,II Regions}

\author{Cintia Quireza\altaffilmark{1,2,3}, Robert T. Rood\altaffilmark{3}, 
Dana S. Balser\altaffilmark{4} \& T. M. Bania\altaffilmark{5}}

\altaffiltext{1}{Observat\'orio Nacional, Rua General Jos\'e Cristino 77, 20921-400, Rio de Janeiro, RJ, Brazil.}
\altaffiltext{2}{Instituto de Astronomia, Geof\'{\i}sica e Ci\^encias Atmosf\'ericas (IAG), Universidade de S\~ao Paulo, Rua do Mat\~ao 1226, 05508-900, S\~ao Paulo, SP, Brazil.}
\altaffiltext{3}{Astronomy Department, University of Virginia,
P.O.Box 3818, Charlottesville VA 22903-0818, USA.}
\altaffiltext{4}{National Radio Astronomy Observatory,
P.O. Box 2, Green Bank WV 24944, USA.}
\altaffiltext{5}{Institute for Astrophysical Research,
725 Commonwealth Avenue, Boston University, Boston MA 02215, USA.}

\begin{abstract}

We report radio recombination line (RRL) and continuum observations of
a sample of 106 Galactic \hii\ regions made with the NRAO 140 Foot
radio telescope in Green Bank, WV. We believe this to be the most
sensitive RRL survey ever made for a sample this large. Most of our
source integration times range between 6 and 90 hours which yield
typical r.m.s. noise levels $\sim 1.0$--3.5 milliKelvins. Our data
result from two different experiments performed, calibrated, and
analyzed in similar ways. A \cii\ survey was made at 3.5 cm wavelength
to obtain accurate measurements of carbon radio recombination lines.
When combined with atomic (\ci) and molecular (CO) data, these
measurements will constrain the composition, structure, kinematics,
and physical properties of the photodissociation regions that lie on
the edges of \hii\ regions. A second survey was made at 3.5 cm
wavelength to determine the abundance of $^3$He in the interstellar
medium of the Milky Way. Together with measurements of the \hepthree\ 
hyperfine line we get high precision RRL parameters for H, \hefour,
and C. Here we discuss significant improvements in
these data, with both longer integrations and newly observed
sources.

\end{abstract}


\section{INTRODUCTION}

In molecular clouds, newly formed massive stars are strong emitters of
ultraviolet radiation. Massive, early-type stars (O4--B9), emit
extreme-ultraviolet ($h\nu >13.6$ eV) photons that will be absorbed by
H atoms creating a bounded zone of H$^+$ that is called an \hii\
region. Those photons which are not energetic enough to ionize
hydrogen and create \hii\ regions penetrate into the surrounding
molecular gas where they can dissociate most molecules and ionize
atoms with ionization potential below 13.6 eV, such as carbon (11.3
eV), silicon (8.2 eV), and sulfur (10.4 eV).  This produces a
photodissociation region (PDR), a neutral hydrogen transition layer
between the relatively hot ($\sim 10^4$ K) ionized gas in the \hii\
region and the cold ($\sim 10$--$100$ K) gas in the molecular
cloud. As the FUV radiation is attenuated across this interface
region, the PDR changes chemical composition from totally ionized to
totally molecular. The processes of ionization, dissociation, and
recombination that occur in a PDR determine its composition, structure
and kinematics.

Any astrophysical environment associated with sources of FUV
radiation should have a PDR. In addition to giant molecular clouds,
examples include reflection nebulae, protoplanetary disks, the neutral 
gas around planetary nebulae, photodissociated winds from red giant and 
asymptotic giant branch stars, the interstellar medium in nuclei of star 
burst galaxies and galaxies with active Galactic nuclei. To date, however, 
most PDR studies have dealt with the gas that lies at the outer boundaries 
of dense, luminous \hii\ regions in the Galaxy (see Hollenbach \& Tielens 1999
and references therein for a comprehensive review about PDRs). The physical 
and chemical processes that regulate dense PDRs are identical to those that 
occur in the cold and warm neutral gas components of the interstellar medium
(see Kulkarni \& Heiles 1987; Dickey \& Lockman 1990; Wolfire et al. 2003).

Observations of a single species in different physical states (ionized, 
atomic, and molecular) may be used to investigate the effects of FUV 
radiation on the structure, chemistry, thermal balance, and evolution 
of the PDR gas. Because the FUV radiation comes from newly formed 
stars, the study of PDRs probes the process of star formation. In particular, 
carbon may be used to derive the physical properties of Galactic PDRs. Carbon 
is the most abundant element with ionization potential lower than that of hydrogen. 
Its dominant molecular form is CO which has been used extensively as a
tracer of molecular gas and regions of star formation (e.g., Dickel et al. 
1978; Viala et al. 1988; Keene et al. 1998; Allen et al. 2004; 
Kulesa et al. 2005; and Jackson et al 2004. See also Jackson et al. 2006 for
a recent survey of $^{13}$CO). 

The fine-structure transition of [\cii] at 158 \micron\ and the recombination 
lines of \cii\ in the radio (e.g. C\,91\,$\alpha$) may be used to 
investigate carbon's singly ionized stage (e.g., Russell et al. 1980; 
Howe et al. 1991; Stacey et al. 1993; Sorochenko \& Tsivilev 2000; 
Lehner et al. 2004). Finally, neutral carbon may be studied using 
the fine structure transitions of [\ci] at 609 \micron\ and 370 \micron\  
(e.g., Phillips \& Huggins 1981; Keene et al. 1985; Genzel 
et al. 1988; Huang et al. 1999; Jenkins \& Tripp 2001; Papadopoulos et al. 2004).

In this paper we discuss the results of an X-band (8-10 GHz)
survey of carbon radio recombination line (RRL) emission from 66
Galactic \hii\ regions. The motivation for this survey was to study
more systematically the \cii\ gas in PDRs. We also present \cii\ RRLs for 
a sample of 47 Galactic \hii\ regions which are part of a program to
measure the abundance of $^3$He in the Milky Way interstellar
medium. These data complement a series of papers (Rood et al. 1984;
Bania et al. 1987, 1997; and Balser et al. 1994), in which we 
reported the observational status of the $^3$He experiment. All
observations were made with the National Radio Astronomy Observatory
\footnote{\nraoblurb } (NRAO) 140 Foot radio telescope in Green Bank, WV. 

We intend to combine our carbon RRL observations with 158 \micron\
transition data from the ISO (Infrared Space Observatory)
archives. Because \cii\ RRLs have a different dependence on the
density and temperature than the [\cii] 158 $\mu$m far-infrared
fine-structure line, measurements of both of these transitions provide
an unambiguous way to derive the physical properties of the \cii\ gas.
This was done by Natta et al. (1994), who observed the C\,91\,$\alpha$
RRL toward the Orion\,A \hii\ region. They suggested that the
C\,91\,$\alpha$ line originates in a dense ($10^6$ cm$^{-3}$) and warm
(500--1000 K) gas, but their estimate still needs to be confirmed with
more high-quality radio line data. Shah (1995) applied the same
technique to W3, NGC\,6334 and W51; his results indicated that the
carbon RRLs form in the same region as the 158 $\mu$m
emission. Finally, in a recent study of carbon RRLs in ultracompact
\hii\ regions (UCHs) Roshi et al. (2005) find that the PDR density is
typically $\gtrsim$\,\nexpo{5}{5}\percc\ and that these UCHs are
embedded in regions of high ambient pressure.

Together with the \cii\ RRLs, we also obtained high precision
recombination lines of H and \hefour\ (henceforth He) as well as radio
continuum intensities for all the Galactic \hii\ regions in our
surveys. These data allow the determination of several physical
parameters, such as the electron density, emission measure, H-ionizing
luminosity, and mass of \hii. Furthermore, we can derive nebular
electron temperatures that can be converted to oxygen abundances
(Shaver et al. 1983). When combined with accurate distance
measurements, these data yield a reliable determination of the
electron temperature and oxygen abundance gradients across the Milky
Way's disk, providing strong constraints on models for Galactic
chemical evolution. The discussion of the physical properties of our
Galactic disk \hii\ regions as well as the temperature/abundance
radial gradients analysis will be presented in future papers.

Here we restrict ourselves to a description of the data and compile
for our source sample the measured properties of the continuum
emission together with the RRL transitions of H, He, and C.  The
spectral line and continuum observations are summarized in \S 2, where
we describe our observing techniques and data reduction process. We
also compile the observed properties of recombination line and
continuum emission for our Galactic \hii\ region sample. In \S 3 we
discuss the overall quality of the data and describe 
its astrophysical potential. A brief summary follows in \S 4.

\section{OBSERVATIONAL TECHNIQUES AND DATA REDUCTION} \label{sec:obs}

\subsection{Radio Recombination Lines} \label{sec:RRL}

\subsubsection{\cii\ survey} \label{sec:cii}

Our potential target list included: (i) all major \hii\ regions
visible to the 140 Foot telescope; (ii) sources that have CO, \ci\,
and \cii\ data; and (iii) sources with embedded B stars as revealed in
IRAS HIRES images at 60 and 100 $\mu$m. Our base sample consisted of
$\sim 100$ sources in the Milky Way visible from Green Bank with
measured X-band Hn$\alpha$ line intensities exceeding 100 mK (Lockman
1989). Our goal was to achieve a spectral r.m.s. noise level of $\sim
2$ mK. We selected 66 objects distributed throughout the Galactic
disk. The majority of these \hii\ regions are located in the first and
fourth Galactic quadrants and their Galactocentric distances can reach
beyond $\sim 10$ kpc.  At 8.6 GHz the $3\farcm20$ HPBW of the 140 Foot
telescope is comparable to that of the CO, \ci, and \cii\ data.
Modeling by Roshi et al. (2005) shows that carbon RL emission near 8.5
GHz is dominated by stimulated emission, and hence we preferentially
observe the PDR material that is in front of the \hii\ region thermal
continuum.

Our spectral line observations were made with the NRAO 140 Foot
radio telescope during the following periods: September and December
1996; January, August, and September 1997; and January, June, and July
1998. The different observing epochs were scattered over the sideral year,
and each source was observed in multiple epochs. We do this to average
out the baseline structure which is a function of sky frequency and
thus changes with the time of the year.\footnote{The same is true of
the \hethree\ experiment.} Detailed discussion of the instrumental 
frequency structure in spectral
baselines can be found in a series of papers by Balser et al.
(1994, 1997) and Bania et al. (1987, 1997). Basically, systematic 
effects produce nonrandom frequency structure in the instrumental
baseline and dominate the radiometer noise at low levels.  These
systematic effects ultimately limit the detectability of weak and
wide spectral lines.

Spectra were taken using the total power position switching observing
mode, where a reference position (OFF position) was observed offset
$\sim 6$ minutes in right ascension from the target source, and then
the telescope was pointed in the direction of the source (ON
position). The integration time on the reference position and the
target position was 6 minutes each for a total of 12 minutes. Where
possible the pointing was determined by peaking on the source
continuum radiation. Otherwise local pointing corrections were
determined using a nearby pointing calibrator. The measured
r.m.s. pointing accuracy was typically $\sim$\,20\arcsec.
Receiver-to-receiver and inter epoch calibration was established with
continuum observations of NGC\,7027 which was assumed to have a flux
density of 6 Jy at 8.6 GHz (Peng et al. 2000). For each source we averaged
all spectra measured during the different observing periods. The resulting 
integration time is the sum of integration times in each of these periods 
and depends on the amount of data, which varies from source to source.

We used the NRAO model IV autocorrelation spectrometer for all the
observations. The autocorrelator was divided into four quadrants, each
sampling a 10 MHz bandwidth with 256 channels which gives a spectral
resolution of $\sim 1.4$ \kms\ per channel. Two autocorrelator
quadrants sampled the 91$\alpha$ band and the other two sampled the
92$\alpha$ band. The pairs of quadrants sampled left and right
circular polarizations. Since recombination lines of the same order
with similar principal quantum numbers, such as C\,91$\alpha$ and
C\,92$\alpha$, should have basically the same intensity, we could
average the C\,91$\alpha$ and C\,92$\alpha$ spectra to attain higher
sensitivity. To calibrate 92$\alpha$ relative to 91$\alpha$ we assumed
that the intensity of the H\,92$\alpha$ transition was the same as the
intensity of H\,91$\alpha$ in the bright \hii\ region W3.

The velocity resolutions of the different frequency bands were not the
same since they were centered at different sky frequencies.  Moreover,
the spectra in different frequency bands are not sampled at the same
velocity.  Therefore it was necessary to regrid one frequency band
onto the scale of the other before averaging the 91$\alpha$ and
92$\alpha$ spectra.  Several experiments with interpolation schemes
gave results in which the r.m.s. noise in the averaged spectrum did
not decrease, i.e., the interpolation introduced noise. We finally
developed a scheme involving shifts in a discrete number of channels
with the size of the shift depending on position in the spectrum. This
was done in a way to give optimum results for the \cii\ line, at the
expense of slightly broadening the H lines. Using this procedure the
r.m.s. noise in the average dropped by $\sqrt{2}$ as it should.

\begin{figure}
\epsscale{0.75}
\plotone{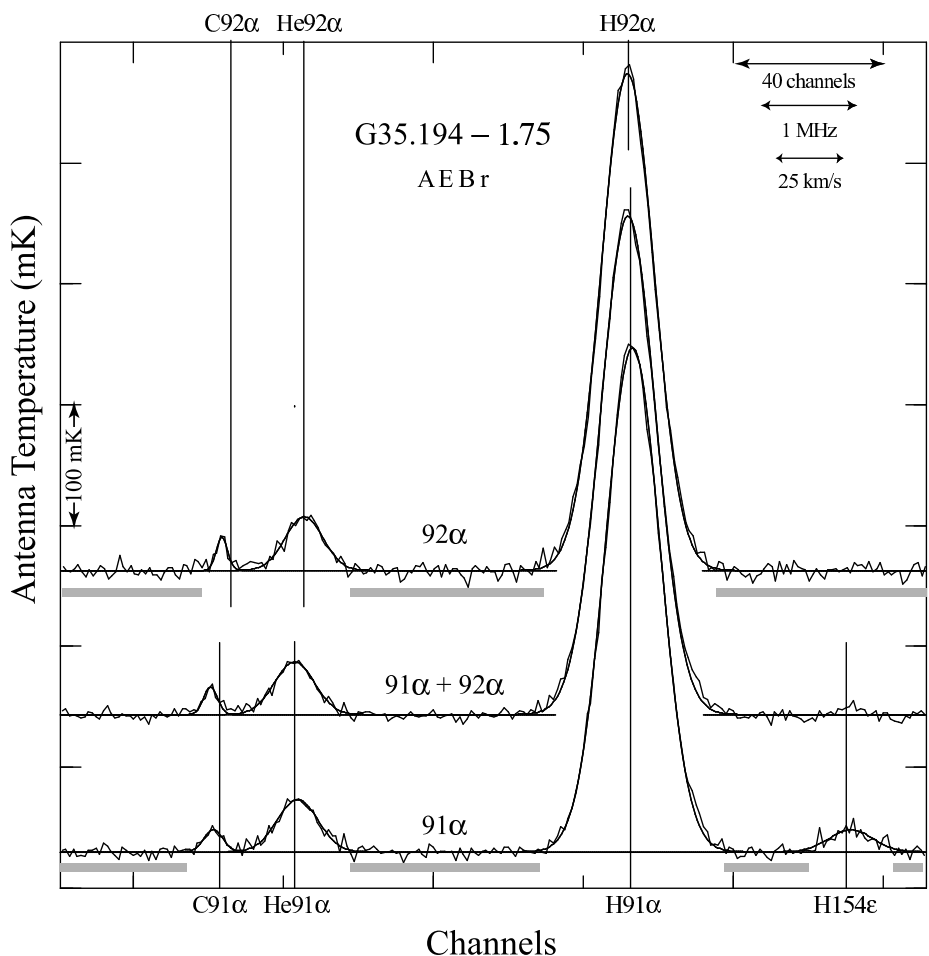}
\caption{Spectra for the Galactic \hii\ region G35.194$-$1.75.  Shown
are the 91$\alpha$ (lower spectrum) and 92$\alpha$ (upper spectrum)
transitions together with the average spectrum: \halpha\ (center
spectrum).  The vertical lines flag the expected line center
positions, and the gray bars denote the channel regions used for the
baseline fit.  A baseline was removed from all these spectra.
Gaussian fits to the lines are also shown. The
H\,154$\epsilon$ line is not used for the baseline fit to the \halpha\
spectrum. The four letters under the source name give the overall
quality factor (QF = A), the reliability factor (RF = E), the C line
quality factor (QF(C) = B), and the C line morphology (r or
resolved). See the text for further explanation.}
\label{fig:aver}
\end{figure}

The process of averaging the two frequency bands is illustrated in
Figure~\ref{fig:aver}, which shows the 91$\alpha$, 92$\alpha$, and the
averaged spectrum (henceforth labeled \halpha) for the \hii\ region
G35.194$-$1.75. Polynomial baseline models were removed from all the
spectra. The gray bars indicate the regions used for the baseline fits;
the \halpha\ baseline regions were the same as for 91$\alpha$. The
\halpha\ spectrum was formed from 91$\alpha$ and 92$\alpha$ spectra 
before any baseline was removed.  The vertical lines show the expected
line centers of the H, He, and C lines. For the 91$\alpha$ spectrum an
additional vertical flag marks the position of the 154$\epsilon$
($\Delta n=5$) transition of H.  To be consistent, we used the flag
marks from the 91$\alpha$ spectrum for the \halpha\ average
data. Gaussian fits to the spectral features are also shown.  The
H\,154$\epsilon$ transition region was excluded from the baseline fits
in both the 91$\alpha$ and \halpha\ spectra. For sources where the
154$\epsilon$ line was weaker than the r.m.s. (it is typically 4--5\%
the H\,91$\alpha$ intensity), we fitted the baseline through that
region.

The regions used to determine the spectral baseline must exclude the
frequencies near the H\,154$\epsilon$ line in the 91$\alpha$ and
the averaged spectrum, while the 92$\alpha$ spectrum has no similar
line.  This could conceivably introduce a systematic error, especially
in the H lines. For better control of the quality of the measured
parameters of the Hn$\alpha$ line we decided to analyze the
91$\alpha$, 92$\alpha$, and \halpha\ spectra individually.

We fitted and removed a 6th-order polynomial baseline model through
regions in which no lines were expected. (This is equivalent to the
12th-order fit in the 20 MHz AC band of the \hethree\ experiment. See
\S\,\ref{sec:He3}.) 
Baselines of this order are {\it required} to model adequately the
{\it real} baseline structure produced by the reflections of source
continuum radiation off the telescope feedlegs and
superstructure. Without such fits one does not recover the expected
r.m.s. from the radiometer equation. (For a complete discussion see
Rood et al. 1984; Balser et al. 1994; 1997; and Bania et al. 1987.)

The line intensities and full width at half-maximum (FWHM) linewidths
were derived using least-squares Gaussian fits. Because they come from
different gas components, the C (PDR) and He (\hii\ region) spectral
features can have a variety of morphologies.  The C line is in fact
often blended with the He line. The worst case appears when the wings
of the He line cover the C line shape producing a composite spectral
feature which can appear as a single very wide line. In many cases the
signal to noise ratio of the data is so large that both the C and He
RRLs can be measured even when the two are blended.

\begin{figure}
\epsscale{0.75}
\plotone{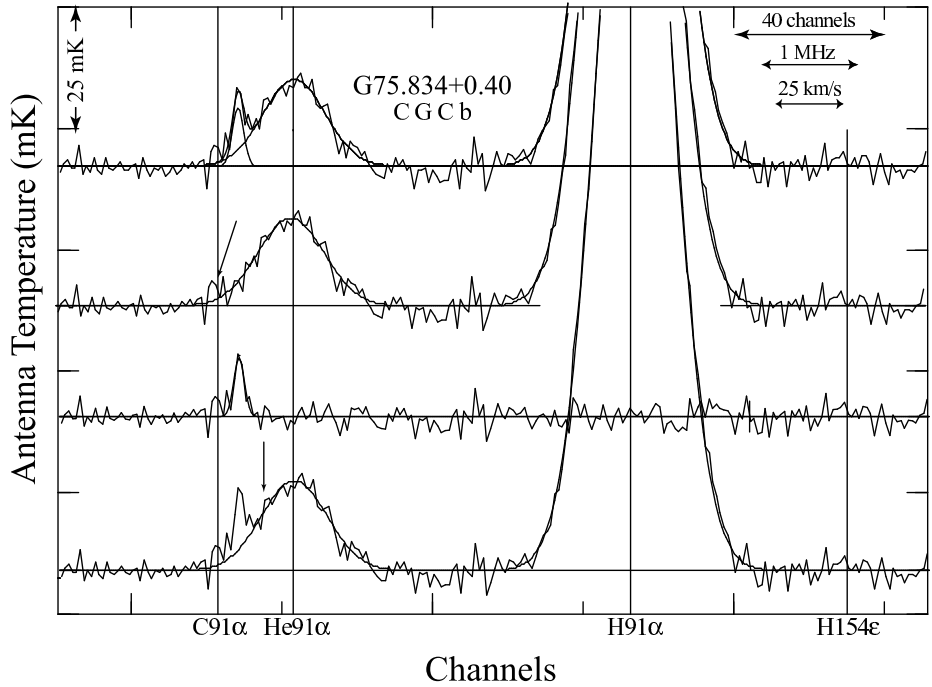}
\caption{An example showing how blended C line parameters are
determined.  The lower plot shows the spectrum for the \hii\ region
G75.834$+$0.40 after a baseline has been removed. The Gaussian has
been fit to that part of the He line to the right of the
arrow---avoiding the C line. The second plot shows the spectrum with
the fit subtracted. A Gaussian fit to the C line is then subtracted
from the original spectrum yielding the third spectrum. The Gaussian
fit to this spectrum extends through the full He line as indicated by
the arrow and shows that the He line is symmetric. Finally the full
spectrum and fits are shown in the upper plot.}
\label{fig:decon}
\end{figure}

We developed an analysis procedure that can extract accurate He and C
RRL line parameters even when the two spectral features are blended.
It is based on the fact that the He emission is generally more intense
than C.  
This procedure is illustrated in
Figure~\ref{fig:decon}. We assume that the He line is symmetric, and first
fit Gaussians to the H line and the part of the He line profile
nearest the H line which is also clearly separated from any C
feature. The arrow in the lower spectrum of Figure~\ref{fig:decon} shows
the left boundary of the fitting region. We then subtract these fits
from the original spectrum which yields a spectrum with only the C
line.  We then make a Gaussian fit to the C line shown in the second
spectrum in order to determine its parameters.  This Gaussian is then
subtracted from the original spectrum. The H and He lines in the
resulting spectrum are then remeasured with the fitting region now
covering the full region of the He line as indicated by the arrow in
the third spectrum. The resulting fit is shown in the upper
spectrum. In principle we could iterate this procedure, but we find
that it is not necessary.

The errors quoted for line intensity and line width are the $\pm 1\,
\sigma$ uncertainty of the final Gaussian fits.  The relative size of
this fitting error generally increases as the line intensity
decreases.  Typical uncertainties in the line intensity and width are:
1\% in both parameters for the H lines; 4\% and 7\% for the He lines;
and 18\% and 25\% for the C lines.  For a small number of sources the
He line shape was not well defined and the Gaussian fit was not
satisfactory.  For these cases the parameters of the He line were
obtained by refitting the Gaussian after fixing the He line center and
width.  The line center was based on the H velocity and the width was
set using the measured H line width and adopting the mean value of the
He/H line width ratio. For these special cases we used the median
value of all the fitting errors for spectral features with similar
intensities.

The baseline removal and the calibration and regridding required to
average the two transitions introduces the potential for some
systematic error. We performed many tests to determine how large this
was. Because of the large sample and high quality of the data, we
could see subtle effects. The effect of the systematic error on the
parameter values for individual \hii\ regions is, however, always
comparable to or smaller than the random error due to radiometer
noise.

In order to further assess the quality of our measurement of the line
intensities and widths, we visually inspected all the data and defined
quality factors (QFs) for the spectrum as a whole and each transition.
The QFs are a subjective rating based on the signal-to-noise ratio,
the structure of the baseline, the crowding of spectral lines, and the
accuracy of the Gaussian fit to the line shape.  Quality {\it A}
sources are our best spectra and quality {\it E} sources are our worse
cases. We have not yielded to the grade inflation so often found in
educational institutions.  Some of our spectra are very, very
good---sometimes the spectra have so little noise they cannot be
distinguished from the Gaussian fits. A ${\rm QF = A}$ meets a very
high standard.  The QFs for the overall spectrum and each of the lines
are relative within each category, so QF(C) = B for the C line does
not necessarily mean the same thing as QF = B for the entire spectrum,
beyond the fact that both are in the second grouping. In all cases
spectra with QFs of D and E give low confidence line parameter
determinations and should be used with caution.

There is an additional factor affecting the C lines. In measuring weak
He lines we are aided by knowing its expected center frequency to high
precision and its intensity and FWHM to a factor of two before we try
to make a fit. This is not the case for the C lines---their precise
positions are not known and anticipated intensities range from zero to
stronger than the He line. C lines should be narrow, but as we will
see below, the widths are highly variable. The narrowness makes it
more difficult to distinguish C lines from noise features. Because of
these factors the details of how the baseline and line profile are
modeled can affect the existence or not of C lines.  For this reason
we introduce a ``reliability factor'' (RF) as a measure of how
confident we are of the reality of a C line. These are designated
excellent, good, fair, poor (E, G, F, P). Lines rated E or G are
almost certainly real. In sources with RF = F there is a small chance
that the C line is not real. For sources with RF = P there is a fairly
good chance that the reported line is not real. In our discussion
below we exclude RF = P lines. The QF we introduced earlier refers to
the quality of the line if it is real.

\subsubsection{$^3$He survey} \label{sec:He3}

This survey is part of an on-going program to measure the abundance of
$^3$He in the Milky Way interstellar medium. Our progress is described
in a series of papers (Rood et al. 1984; Bania et al. 1987; 1997;
2002; and Balser et al. 1994). As part of this effort we obtain very
sensitive measurements of a number of recombination lines in Galactic
\hii\ regions.  Observations were made with the NRAO 140 Foot
telescope during different epochs from February 1982 through July,
1999.  At first we observed weak H and He recombination lines
primarily for the purpose of characterizing the baseline frequency
structure, monitoring the system performance, and developing data
reduction procedures. By the mid 1980's, however, it became clear that
the recombination lines were a useful tool for probing the structure
of \hii\ regions, so we established a uniform intensity scale
calibration for all the recombination lines we observed in the entire
survey (Bania et al. 1997).

Here we report new observations of RRLs that represent substantial
improvements over our previously reported measurements: (1) we added
more integration time for some sources; and (2) we increased the number
of \hii\ regions in the original sample by 25 objects.  Because of this
the line parameters we report for some sources are different than
previously published.  In no case do these new line parameters exceed
our published errors. 
{\it The improved data reported here supersede all our
previously reported recombination line measurements for these objects.\/}

Our current source sample includes 47 \hii\ regions. Our targets are
sources with reasonably well determined distances and they are well
distributed throughout the Galactic disk.  The selection criteria
varied over time as we realized the optimum source characteristics for
detecting the very weak \hepthree\ line. Low density, large angular
size, distant sources are best. Many classic \hii\ regions such as
Orion\,A and M17S are dreadful \hepthree\ targets when observed by
traditional design radio telescopes like the 140 Foot. The integration
times are quite large, so the source sample is smaller in number than
the \cii\ survey. The velocity resolution is a factor of 2 larger than
the \cii\ survey so the C lines are more likely to be unresolved in
velocity width. On the other hand, no other survey reaches such low
levels so the data are an important complement to the \cii\ survey. As
in the \cii\ survey, most of the \hii\ regions of the $^3$He survey
are located in the first and fourth Galactic quadrants. Because we
were searching for a \hethree\ abundance gradient, the survey also has
many more sources with galactocentric distances greater than 8.5 kpc
with a few as far out as $\sim 19$ kpc.  A few sources are located at
the far side of the Galactic Center.

All the recombination line observations reported here were made with
the NRAO 140 Foot telescope which has a HPBW of $\sim\,3\farcm20$ at
these wavelengths.  Our new observations were made during 17 observing
sessions spanning the period between 1996 July and 1999 July. These
new data were obtained and analyzed in the same way as described by
Rood et al. (1984), Bania et al. (1987, 1997), and Balser et
al. (1994). We used the same procedures as described in
\S\,\ref{sec:cii} to acquire the spectra. The autocorrelator was
divided into four quadrants, each sampling 20 MHz wide sections of 256
channels each, providing a spectral resolution of $\sim 2.7$\kms. We
observed left and right circular polarizations of the $^3$He$^+$ line
in two quadrants and various recombination lines in the other two.

Here we report the results for only the 91$\alpha$ spectra.  This
spectral band sampled the C\,91$\alpha$, He\,91$\alpha$,
H\,91$\alpha$, and H\,154$\epsilon$ RRL transitions.  For each source
we averaged all spectra measured during and following August 1992.
Because uniformity is an important characteristic of the results
presented here we chose not to include data before this period when
our observing and data analysis procedures were different.  Therefore,
in addition to the new epochs listed above, we included the
observations made between August 1992 and March 1996 which are
reported in Bania et al. 1997 (epochs 8--12 in that paper).  The
amount of new data varies from source to source. Some objects such as
W3 were observed during all observing periods. Others, Orion\,A for
example, have a small amount of data because they were observed only
for system tests.

Typical spectra are shown in Bania et al. (1997).  A 12th-order
polynomial baseline model was removed from each averaged
spectrum. The spectral bandwidths for the $^3$He and
\cii\ surveys are 20 and 10 MHz, respectively.  
For this reason the order of the polynomial used in the $^3$He survey 
is double that of the polynomial used in the \cii\ survey. 
Finally, we assigned QFs for the $^3$He survey spectra using the same
criteria described in \S\,\ref{sec:cii}.  

\subsubsection{Radio Recombination Line Properties}\label{sec:combined}

Table~\ref{tab:properties} compiles properties for our complete sample
of Galactic \hii\ regions which combines the \cii\ and $^3$He
sources. Listed are the Galactic and equatorial (epoch 1950.0)
coordinates, the total integration time, the r.m.s. noise in the 
spectral baseline region and the QF for the spectra.  The last 
column of Table~\ref{tab:properties} identifies the source's survey
membership.  The two surveys have 7 objects in common. Also, five
extended, morphologically complex \hii\ regions were observed at two
or three different positions. Specifically, these sources are M16
(M16 and M16N), M17 (M17N and M17S), NGC\,6334 (NGC\,6334-A
and NGC\,6334-D), Rosette (Rosette-A and Rosette-B), and S209
(S209, S209N and S209S). Altogether we have taken spectra toward 
119 directions for a sample of 106 distinct HII region complexes. 

Tables~\ref{tab:H}, \ref{tab:He} and \ref{tab:cii} summarize the
observed properties of the H, He, and C RRLs, respectively. Listed
for each source are the peak intensity ($T_{\rm L}$ in milliKelvins of
antenna temperature), the full width at half maximum line width
($\Delta v$ in \kms), and the radial velocity relative to the Local Standard 
of Rest \footnote{The 140 Foot velocities cited here are in the kinematic 
LSR frame using the radio definition of the Doppler shift. The kinematic 
LSR is defined by a Solar motion of 20.0 \kms\ towards ($\alpha$,$\delta$)
= (${\rm 18^{\rm h}}$,+30\degr)[1900.0].} ($V_{LSR}$ in \kms) measured at 
line center along with their associated standard deviations. 
Velocities in the 6 \& 7th columns of table~\ref{tab:H} 
correspond to the LSR velocity and the $1\,\sigma$ error, 
assumed for the source, respectively.
Also listed is the quality factor (QF) for the line measurement.  

\begin{figure}
\epsscale{0.75}
\plotone{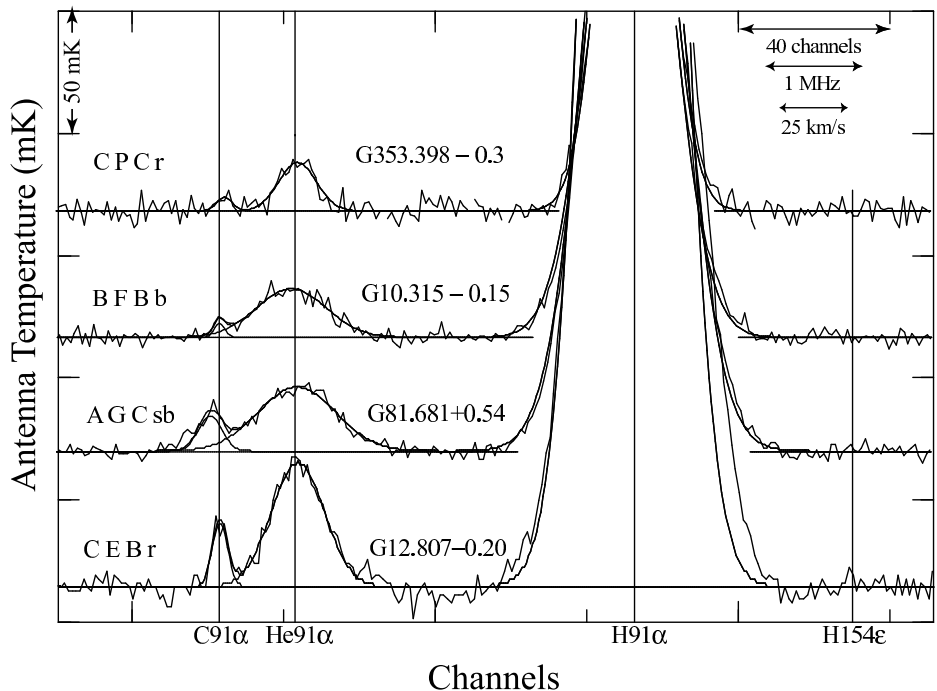}
\caption{Examples from the \cii\ survey of the morphological
classification of C\,91$\alpha$ RRL emission.  Spectra with differing
quality factors (QF \& QF(C)) and reliability factors are shown. Each
spectrum is labeled from left to right with QF, RF, QF(C), and 
morphology. The \hii\ region G12.807$-$0.20 has a carbon recombination
line emission line classified as resolved (r).  The C line in
G81.681+0.54 is classified as semi-blended since it occurs in the
wings of the very wide He line. Both of these are solid detections
with reasonably well determined line parameters. In contrast the
blended (b) C line in G10.315$-$0.15 and resolved C line in
G353.398$-$0.3 are weak and their reality might be debated.}
\label{fig:cii-ex}
\end{figure}

\begin{figure}
\epsscale{0.75}
\plotone{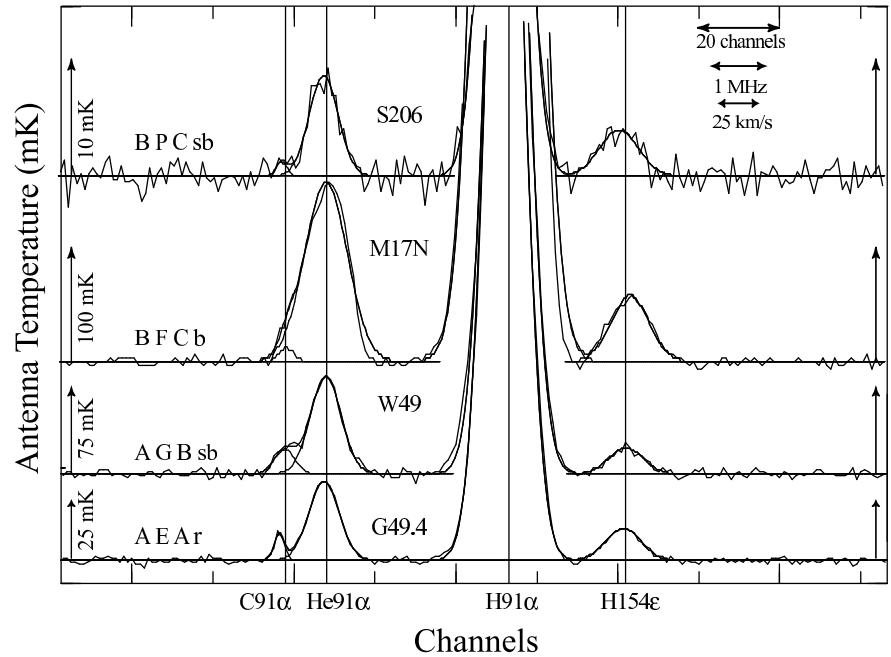}
\caption{Examples from the \he3\ survey of the morphological
classification of C\,91$\alpha$ RRL emission.  The intensity scale
varies from spectrum to spectrum and is indicated by the length of the
arrows at the left of each. Spectra with differing quality factors (QF
\& QF(C)) and reliability factors are shown. Each spectrum is labeled
from left to right with QF, RF, QF(C), and morphology. The \hii\
region G49.4 is in the highest category of each ranking; the C and He
line data are indistinguishable from the Gaussian fit.}
\label{fig:he3-ex}
\end{figure}

Figures~\ref{fig:cii-ex}$-$\ref{fig:he3-ex} show examples of different
types of C RRL spectra that have been obtained in these surveys.
Because the C RRL lineshapes can be quite complex we have also
classified each source's lineshape morphology.  The combination of
spectral crowding, linewidths, and velocity field structure together
produces a variety of composite spectral features. In many cases the C
and He is blended into what appears to be a single very wide line
whose width begins to approach that of the smallest scale instrumental
baseline frequency structure. In other cases these lines are
completely distinct one from the other. We distinguished three
different morphologies for the C lineshape.  The C emission line is
deemed ``resolved'' (r) whenever it is completely distinct from any He
emission. In this case the wings of both lines are not superposed.
The C line morphology is ``semi-blended'' (sb) whenever the C and He
line profiles overlap.  It is nonetheless always possible to identify
the center of both lines.  Finally, the morphology is ``blended'' (b)
whenever the C and He line profiles both overlap and have a linewidth
wherein it is difficult to identify the C line center.  Table~\ref{tab:cii} 
compiles the morphology classification and reliability factor (RF) for
the carbon recombination lines.  

Each spectrum in Figures~\ref{fig:aver}--\ref{fig:he3-ex} is labeled
with 4 letters. From left to right these indicate the overall spectrum
quality factor (QF = A--E), the reliability factor (RF = E, G, F, P),
the C line quality factor (QF(C) = A--E), and the C line morphology
(r, sb, b). We have purposefully chosen some of our lower quality data
to illustrate what we mean by our QF and RF ratings. In
Figures~\ref{fig:cii-ex}\&\ref{fig:he3-ex} there are a few points
worth commenting on:

\begin{itemize}

\item The \hii\ region G49.4 is in the highest category of each
  ranking.  Both the C and He line data are indistinguishable from the
  Gaussian fit.

\item Before deconvolution one might be inclined to think
  that the C line we identified from the M17N spectrum was due to a
  separate velocity component of the He line.   While one sees evidence
  for a separate velocity component on the high velocity side (right)
  of both the H and He lines, there is, however, no evidence for this C 
  component.  The signal to noise ratio is high enough
  that our deconvolution procedure recovers a relatively weak,
  Gaussian shaped C line that could have easily been overlooked.

\item The \hii\ regions G81.681$+$0.54 and G10.315$-$0.15 have, respectively, 
  semi-blended and blended C lines because of the large width of the
  He lines. G75.834$+$0.40 (Figure~\ref{fig:decon}) has a
  ``normal'' He line width and the C line is blended because the C
  line velocity is offset.

\item  Sources with rather similar looking features such as G10.315$-$0.15
  and G353.398$-$0.3 can end up with different RFs.  Additional
  information beyond the final averaged spectrum is used to determine
  the RF.  For example, the 91$\alpha$ and 92$\alpha$ spectra are
  examined separately for consistency.

\item Many of the C lines with RF=P are real. For example,
  Balser (private communication) has obtained a very long integration
  on S206 using the Green Bank Telescope, which shows the C line with
  a very high signal to noise ratio.

\end{itemize}

The 7 objects common to the two surveys provide a means to assess the
accuracy of the measured line parameters. In some cases the two
surveys used slightly different positions for the sources that might
compromise a comparison.  Nevertheless, the spectral line data and
total power continuum (see \S\,\ref{sec:cont}) are in good agreement
for these common sources.  Exceptions occur when the signal to noise
ratio is low.  In some cases discrepancies occur for the weak, narrow
C line which can be more easily affected by a narrow noise feature
than the wider H or He lines of comparable intensity.

We give two measurements of carbon lines for M16 and G23.706+0.17 in
Table~\ref{tab:cii}. The PDRs in our sample have unknown density
structure and velocity fields. We therefore tried to measure every
single signal that might reasonably be a carbon line. In both M16 and
G23.706+0.17 the C lines are very weak; it is possible that neither of
them is real. If real, however, these lines must be produced by
physically distinct components since they appear at different Doppler
shifts. The M16 carbon lines at 25.3\,\kms\, and 38.2\,\kms\, have
intensities of 3.9 and 2.8 mK, respectively, with an r.m.s. of 1.4
mK. (We did not detect the 38.2\,\kms\ line in the \cii\ survey.)  For
G23.706+0.17 the C lines are found at 100.1\,\kms\ and 116.9\,\kms\,;
their respective intensities are 1.5 mK and 3.6 mK with an r.m.s. of
1.0 mK.

\subsection{Radio Continuum} \label{sec:cont}

When combined with RRL data, radio continuum measurements provide
essential information for calculating the physical properties of \hii\
regions.  We therefore made two independent radio continuum emission
surveys with the NRAO 140 Foot radio telescope for our sample of
Galactic \hii\ regions. In both surveys we made cross scans in right
ascension (RA) and declination (DEC) through the center of the
source. Data were obtained for right and left circular polarizations.
The nominal frequency of these continuum data was 8665 MHz and we
typically observed with a 300 MHz bandwidth.  We typically achieved an
r.m.s. sensitivity of 10 mK antenna temperature.  This radiometer
r.m.s. can often be compromised by much larger uncertainties caused by
source confusion and by difficulties in establishing the continuum
background emission which must be subtracted in order to get the true
source continuum.  The flux density scale was determined by continuum
observations of NGC\,7027 which was assumed to have a flux density of
6 Jy (Peng et al. 2000).

The first survey was made during the same epochs as the line data (see
\S\,\ref{sec:cii}\,\&\,\ref{sec:He3}). The continuum measurements were
interleaved with spectral line scans at $\sim 2$ hour intervals.
We normally used the switched power (SP) technique. That is, during
the continuum cross scans the Cassegrain subreflector was nutated
(at $\sim 2.5$ Hz) to an OFF position. The OFF was 
$\sim 8\arcmin$ from the position along the path through the center of the
source. The direction of the beam throw was position angle 202\degper5
(recall that the 140 Foot is an equatorial mounted telescope).

The switched power technique has a distinct advantage over non-nutated
total power (TP) scans. The effect of the atmosphere is more
effectively removed with the SP technique. This is especially
important under marginal weather conditions where one needs to be
especially vigilant about maintaining intensity calibration. The
disadvantage of the SP technique is that for extended sources the OFF
position may still be within the source itself and in crowded parts of
the Galaxy the OFF may include other sources. These effects are
relatively easily identified in the resulting continuum scans. They
also make data analysis difficult because the continuum emission from
the OFF position, whether due to substructure within the \hii\ region
itself or from an ``intruder'' source, is subtracted from the signal
coming from the ON position.

Depending on the intensity and sky position of this unexpected OFF
contribution, the true shape of the source continuum emission
structure in these intensity versus sky position cross scans may be
altered. Emission in the OFF can also corrupt data at the position
extremes of the scan, making it difficult or even impossible to fit a
baseline. These effects can drastically compromise the continuum
measurement.

For these reasons we made a second continuum survey.  The \cii\ survey
objects were observed in March 1999; $^3$He sources were observed in
May 1999. These data were obtained for all sources near transit
(nearly zero hour angle) and under good weather conditions. We were
thus able to observe with the total power (TP) technique.  Since the
subreflector was not nutated, these data do not suffer from any of the
continuum confusion or self-chopping issues describe above for the SP
technique.

In both the total and switched power surveys, RA and DEC scans for
both polarizations were individually analyzed. Polynomial baseline
models were fitted through regions having no continuum emission from
the source.  The order of the polynomial used varied from a linear to
a 3rd-order fit. Higher order fits were used whenever the baseline
regions were sufficiently large. For extended sources and sources in
crowded fields, these baseline regions were sometimes very limited. In
many cases distinguishing between the source continuum emission and
the sky and continuum background level was quite tricky. To aid this
process, we identified the position of each of our target sources in
continuum maps and compared our observations with features in these
maps.  We used continuum maps at 5 GHz (Altenhoff et al. 1979; Haynes
et al. 1978) and 2.7 GHz (F\"urst et al. 1990; Reich et al. 1990).

After removing the polynomial baseline, we measured the continuum
intensity and angular size (full width at half-maximum; FWHM) using a
least-square Gaussian fit to the RA and DEC cross scans. For each scan
a single Gaussian was fit to the central peak; the fit was confined to
an angular range comparable to the telescope beam size.  For SP data
we also averaged over the different observing epochs.  Before
averaging, these scans were checked for quality. Specifically, for
each source we checked for consistency in the fitted parameters for
different scans. Generally the agreement was satisfactory. Scans with
discordant parameters were reanalyzed and excluded in cases where the
inconsistency persisted. These discarded data generally had pointing
problems or bad Gaussian fits.


As we did for the RRL spectra, we also visually inspected our
continuum scans and assigned quality factors to the data. The
continuum QFs are based on the accuracy of the Gaussian fit to the
emission feature, the uncertainty in the baseline models and the
contrast between the target source emission and any emission from
possible sub-structures or nearby confusing sources.  Again, quality
{\it A} sources are our best measurements and quality {\it E} sources
are our worse cases. We deem that sources with QFs of {\it D} and {\it
E} give low confidence continuum emission parameter determinations.

Near the end of our surveys we discovered an offset between the SP and
TP intensity scales. Based on calibration sources NGC 7027 and Virgo A
the SP intensity was $1.15 \pm\ 0.06$ times greater than that of the
TP intensity. This offset had not been noticed by other observers or
NRAO staff. Because the 140 Foot telescope was closed, we were not
able to find the source of this offset. Since the basic calibration of
the telescope for our surveys was made using the SP technique, we used
this correction factor to convert all TP intensities and the
associated errors to the SP intensity scale.

Tables~\ref{tab:conttp} and \ref{tab:contsp} summarize the observed
properties of the continuum emission for our source sample. Listed for
both the TP and SP surveys are the peak continuum intensity in Kelvins
of antenna temperature and the FWHM observed angular size in arcmin
together with the associated errors determined from the Gaussian fits
for both of the RA and DEC scans. TP continuum scans are already
corrected by the 1.15 scale factor just described. The geometric mean
for the average RA and DEC values is also presented. The quoted errors
in the continuum intensity are the Gaussian fitting errors. Due to
baseline problems and complex structures uncertainties in the peak
intensity are certainly larger. Such errors might be 10\%,
or even higher than 20\% in the worst cases.  Since we have no way to
quantify those uncertainties, we crudely estimate this error by 
quality factors (QF) for the continuum data.  These QFs are listed 
separately in Tables~\ref{tab:conttp} and \ref{tab:contsp}.

Any comparison of the continuum observations of the 7 sources in
common between the \cii\ and $^3$He surveys is compromised by the
slightly different positions used and the fact that most of these
sources have complicated structure which appears in the OFF beam of
the SP observations. The best comparison can be made with the TP
observations where we calculate an average intensity ratio between the
$^3$He and \cii\ survey to be $0.97 \pm\ 0.03$.

Finally, we suggest how to best use these continuum data. As a source
was tracked on the sky during these surveys, there were small
corrections for changes in the telescope gain with elevation and also
corrections due to weather. Since the data were taken in an
interleaved manner, these corrections were the same for the average SP
continuum and the line measurements. Thus, when calculating properties
that depend on the line-to-continuum ratio, such as the electron
temperature, we suggest using the SP continuum measurements. On the
other hand, for calculating properties that depend only on the
continuum data we suggest using the TP measurements. These are our
most accurate continuum parameters since they were taken during good
weather conditions at transit where the calibration is most accurate.

\section{Characterization of the Data}\label{sec:characterization}

In a series of papers to follow we will discuss astrophysical results
derived from the data compiled here. Still we feel that it is
important to give some impression of the overall quality of the data
and the sort of conclusions that can eventually be drawn.

\begin{figure}
\epsscale{0.75}
\plotone{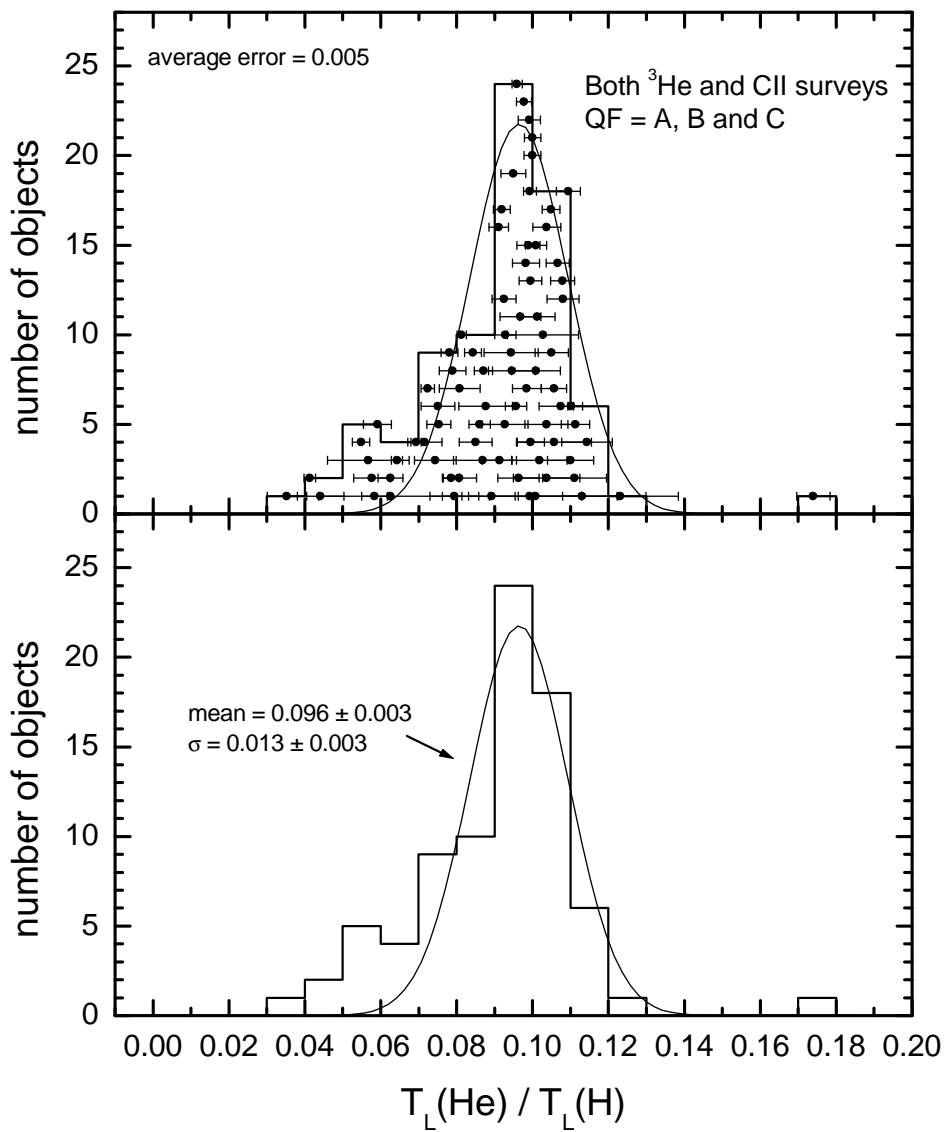}
\caption{The distribution of the ratio of He and H line
intensities. Only the highest quality data (QF = A--C) are used. The
lower plot shows a histogram of the distribution and the best fit
Gaussian. These are repeated in the upper plot now with each data
point shown along with its error bars. Within each histogram bin the
points are plotted with their measured $x$ (in this case $T_L({\rm
He})/T_L({\rm H})$) and the points are added to each bin in order of
increasing $T_L({\rm H})$.
\label{fig:tlHe}}
\end{figure}

Figure~\ref{fig:tlHe} shows the distribution of the ratio of He to H
line intensities. The lower plot shows a histogram of the distribution
and the Gaussian which best fits that histogram.  It is important to
determine whether the width of this distribution is due to an
intrinsic source to source spread or to the errors in individual \hii\
region parameter values.  In order to better illustrate the size of
the errors we construct in the upper panel a pseudo-histogram plotting
each data point along with its error bars. Within each histogram bin
points are added in order of increasing $T_L({\rm H})$; the $n$th
point is plotted at $y = n$ and $x = T_L({\rm He})/T_L({\rm H})$. The
histogram and Gaussian are repeated for reference.

The error bars for the Figure~\ref{fig:tlHe} sources are typically
much narrower than the width of the observed distribution. The average
error is 0.005, less than one half the observed width of
0.013.\footnote{When referring to the widths of the distributions we
quote the value for $\sigma$ in the best Gaussian fit. The FWHM of the
distribution is $2.35\sigma$.} This indicates an intrinsic spread in
$T_L({\rm He})/T_L({\rm H})$. Near the peak and toward higher values
the spread is well modeled as a Gaussian and could be due to
variations in the He abundance, degree of ionization, turbulence, or
other property of the \hii\ region.  There is an obvious skew in the
direction of low values for $T_L({\rm He})/T_L({\rm H})$. This is
probably due to the fact that some of the He is not ionized in these
\hii\ regions. We suspect that there is a real paucity of sources with
$T_L({\rm He})/T_L({\rm H}) < 0.04$. We have high enough sensitivity
that we would have detected such sources unless they are relatively
uncommon. 

\begin{figure}
\epsscale{0.75}
\plotone{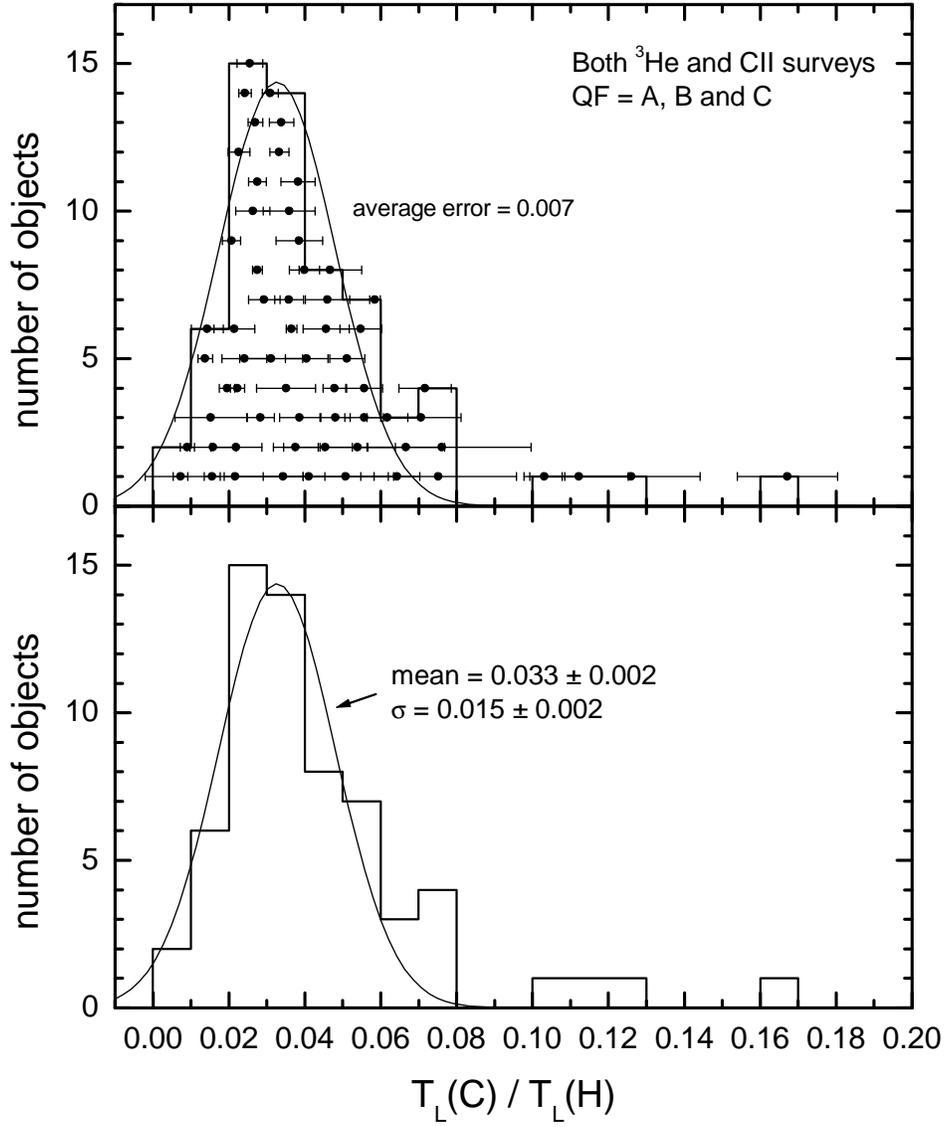}
\caption{The distribution of C line intensities.  See
Figure~\ref{fig:tlHe}.
\label{fig:tlC}}
\end{figure}

Figure~\ref{fig:tlC} shows the analogous distribution for C lines. The
distribution is again much broader than can be attributed to
measurement errors alone. The average error of 0.007 is about half
the observed width. The observed distribution is well modeled as
Gaussian except for a tail toward high values. The drop off toward
small values is most likely due to lack of sensitivity.

\begin{figure}
\epsscale{0.75}
\plotone{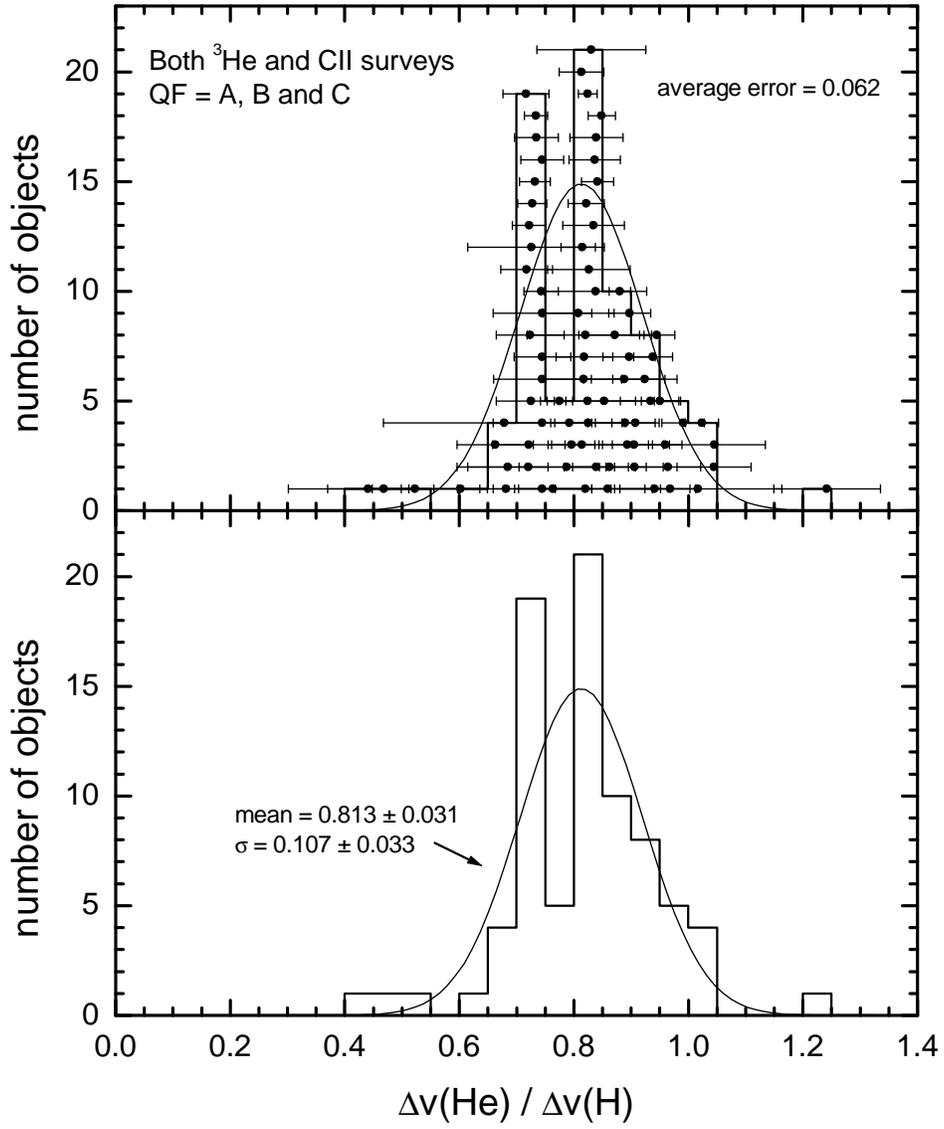}
\caption{The distribution of He line widths.  See Figure~\ref{fig:tlHe}.
\label{fig:dvHe}}
\end{figure}

\begin{figure}
\epsscale{0.75}
\plotone{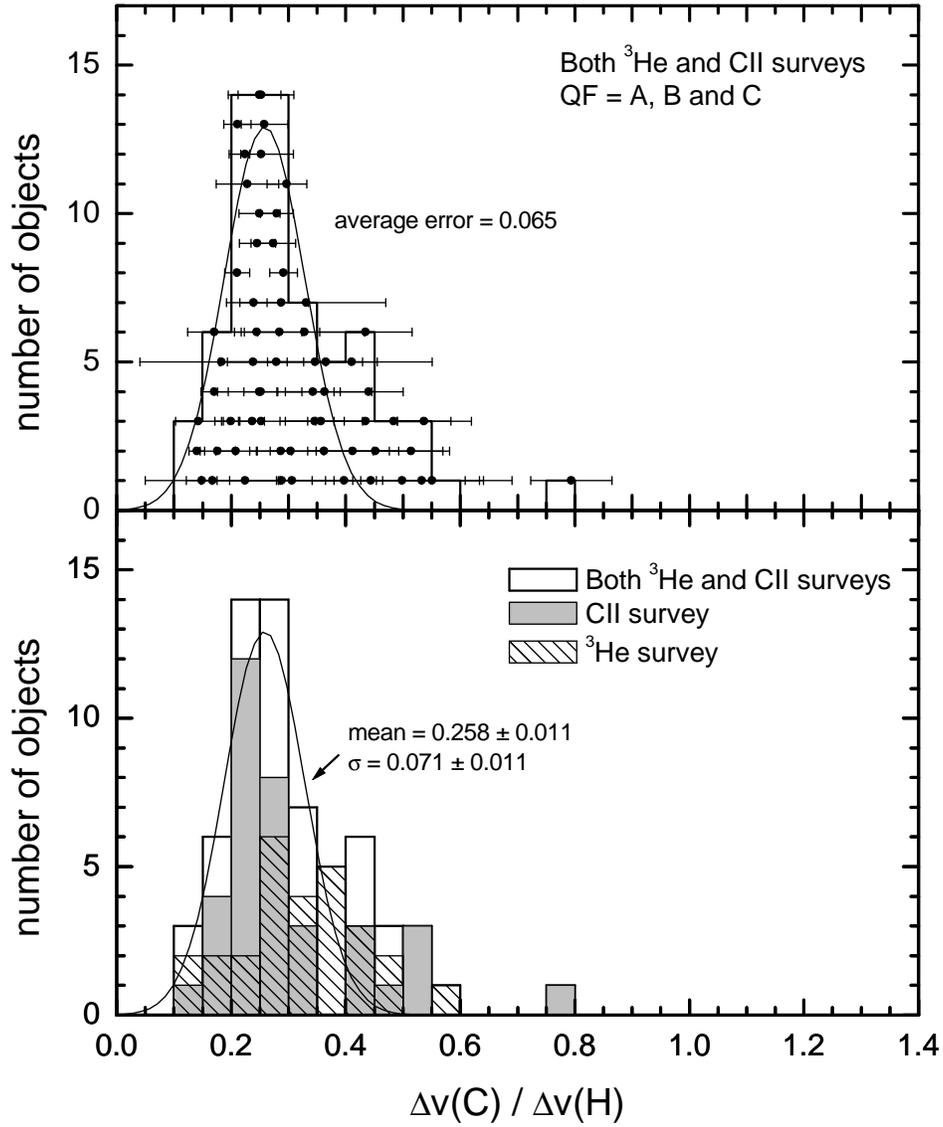}
\caption{The distribution of C line widths for the combined \cii\ and
\hethree\ surveys. In the lower panel the component of the
distributions from the \hethree\ and \cii\ surveys are shown.
See Figure~\ref{fig:tlHe}.
\label{fig:dvC}}
\end{figure}

The line width distributions are shown in Figures~\ref{fig:dvHe} and
\ref{fig:dvC}. Both have substantial intrinsic width, and neither
distribution is well described as a Gaussian. For He the distribution
of $\Delta\,v({\rm He})/\Delta\,v({\rm H})$ drops off sharply for
values $< 0.7$ and $\ga 1$. Line widths arise from thermal broadening
and turbulence. The \hii\ regions G10.315$-$0.15 and G81.681+0.54
(Figure~\ref{fig:cii-ex}) have intrinically wide He lines. In the
absence of turbulence $\Delta\,v({\rm He})/\Delta\,v({\rm H})=
0.5$. \hii\ regions with $\Delta\,v({\rm He})/\Delta\,v({\rm H})$
approaching 1 must either be very highly turbulent or the ionized He
and H are not co-spatial.

Since the C lines arise from the PDR one expects them to be narrower
both because of the lower temperature and the higher mass of C. The
mass effect alone would yield a C line a factor of $\sim 3.5$ more narrow
than H. In the \cii\ survey the most likely line width is 0.24 the
mean H line width or roughly 6.2\kms, and in the \hethree\ survey it is
0.31 the mean H line or 7.5\kms. In the combined survey the most
likely width is 0.26 the H line width. The resolution of the two
surveys was respectively 1.4 and 2.7\kms, so the C lines are typically
resolved in velocity. The lower panel of Figure~\ref{fig:dvC} shows
histograms demonstrating how the two surveys independently contribute
to the combined survey.
The higher resolution \cii\ survey does turn up more sources with
narrower lines, but curiously yields no sources with velocity widths
less than the resolution of the \hethree\ survey. 
Both surveys show a scatter of sources with larger
widths, ranging up to $\sim 12\kms$. We do not have the sensitivity to
detect even larger widths.

\begin{figure}
\epsscale{0.75}
\plotone{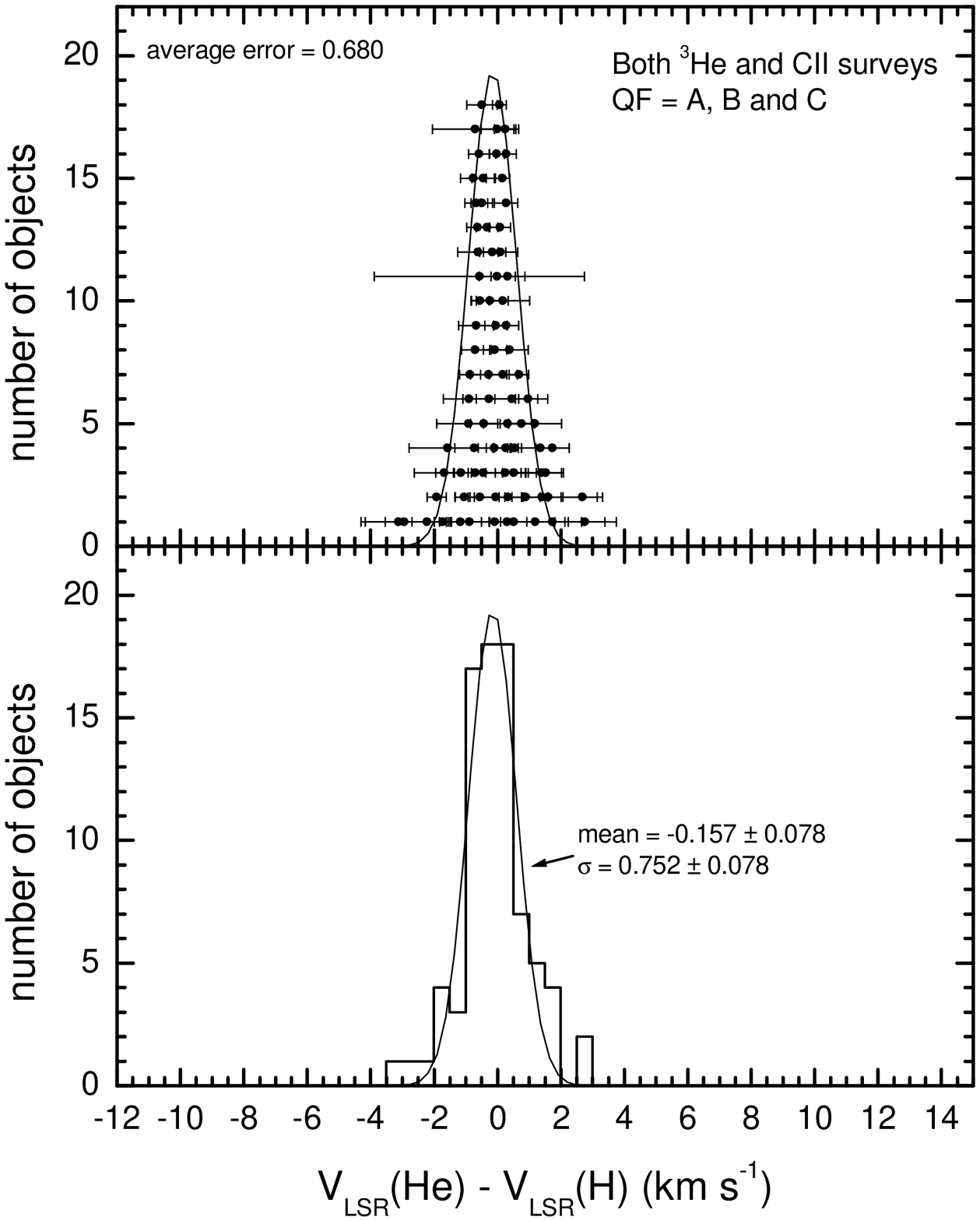}
\caption{The distribution of He line velocities.  See Figure~\ref{fig:tlHe}.
\label{fig:vHe}}
\end{figure}

\begin{figure}
\epsscale{0.75}
\plotone{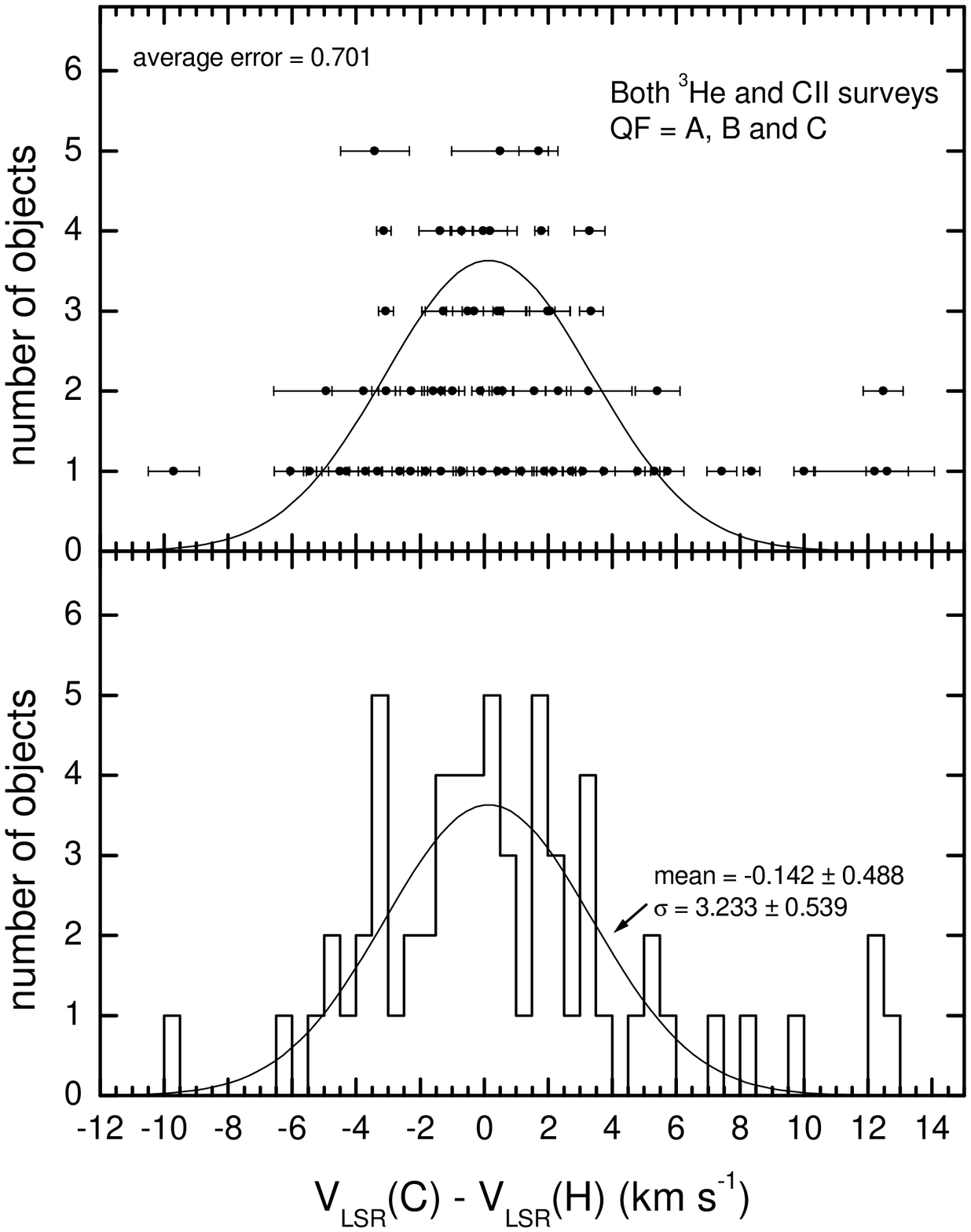}
\caption{The distribution of C line velocities.  See Figure~\ref{fig:tlHe}.
\label{fig:vC}}
\end{figure}

\begin{figure}
\epsscale{0.75}
\plotone{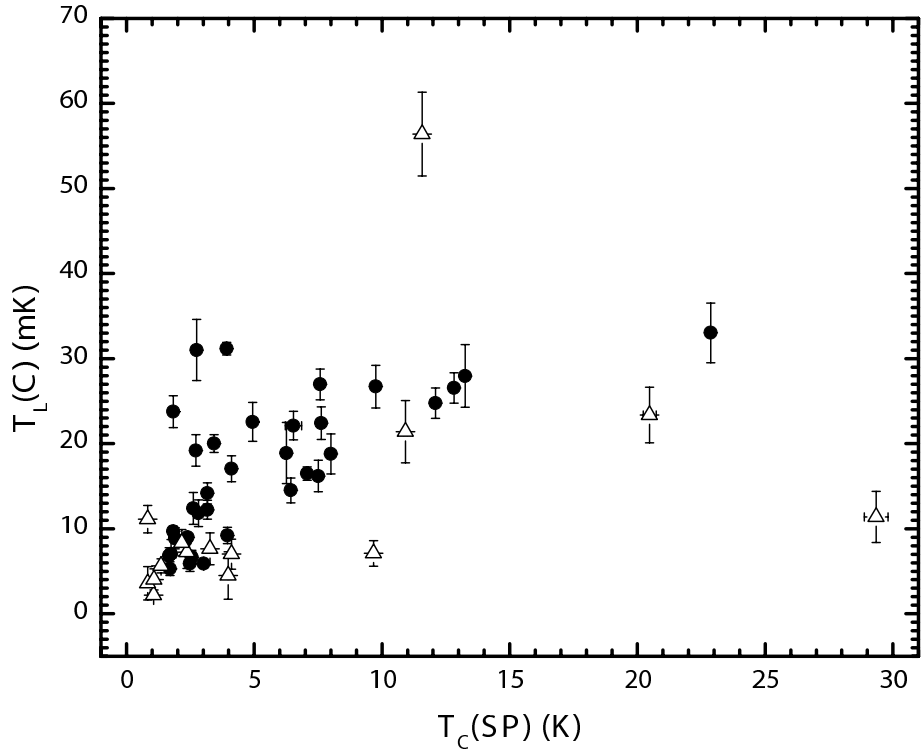}
\caption{Carbon line intensity versus continuum
intensity for sources with best QFs ({\it C} or better) 
for both line spectra and continuum.  The solid circles and open triangles correspond to the
\cii\ and \he3\ surveys, respectively.  A linear least-squares fit
including both surveys yields $T_{\rm L}(C) = (9.52 \pm\ 1.73) + (0.90
\pm\ 0.22)\,T_{\rm C}(SP)$. Only sources with continuum measured using
SP are included.
\label{fig:se}}
\end{figure}

The line velocity distributions relative to the H line are shown in
Figures~\ref{fig:vHe} and \ref{fig:vC}. For He the peak of the
Gaussian fit is near zero and the width of the Gaussian fit is
consistent with that expected from the individual source errors
(Figure~\ref{fig:vHe}).  This suggests that overall the H and He
density and ionization distributions are similar over the $3\farcm20$
telescope beam.  Figure~\ref{fig:vC} shows that the PDR gas, probed by
the C RRLs, is typically moving several \kms\ relative to the \hii\
region, with a few velocities ranging up to 10\kms.  The distribution
of velocities is roughly symmetric around zero with $\sigma =
3.2$\kms. From these data alone we have no way to distinguish between
gas flowing toward or away from the \hii\ regions. We also have no
definitive detection of sources with multiple velocity
components. Naively one might have expected to find a few sources with
PDR's both in front and beyond the \hii\ region.\footnote{In some
cases the angular size of the \hii\ region is larger than the
telescope beam, which might lead us to miss some PDR material with
small radial velocity.} If the carbon emission line is dominated
by stimulated emission, we would predominantly observe PDRs in front
of the \hii\ regions. Figure~\ref{fig:se} shows that the C line
intensity is strongly correlated with the \hii\ region continuum
intensity, which indicates that stimulated emission is
occurring. This is consistent with similar observations toward 11
ultracompact \hii\ regions by Roshi et al. (2005). 

\section{SUMMARY}

We report high precision radio recombination line and continuum
observations made with the NRAO 140 Foot telescope near 8.5 GHz for
106 Galactic \hii\ regions.  Most of our source integration times
range between 6 and 90 hours which yields typical r.m.s. noise levels
$\sim 1.0$--3.5 milliKelvins. These data come from two different
surveys that were observed and analyzed in similar ways.  Our sample
consists of \hii\ regions located throughout the Galactic disk with a
wide range of physical properties. Hydrogen recombination emission was
detected in all 106 \hii\ regions with line intensities 27--4277 mK.
Helium was detected in 77 \hii\ regions with line intensities 3--468
mK.  We detected carbon in 74 objects with intensities 1--200 mK.

\acknowledgments

We thank the staff of NRAO Green Bank for their help, support and
friendship. The \he3\ research has been sporadically supported by the
National Science Foundation.  The most recent grant was AST 00-98047.
CQ is grateful to the Department of Astronomy at the University of
Virginia, for their hospitality. Her work was partly supported by
the Levinson Fund of the Peninsula Community Foundation and FAPESP.


\clearpage
\begin{deluxetable}{llrrrrrrrr}
\tabletypesize{\scriptsize}
\tablecaption{Properties of Galactic \hii\  Regions}
\tablewidth{0pt}
\tablehead{
\colhead{Source} & \colhead{Name} & \colhead{l} & \colhead{b} & \colhead{RA(1950.0)} & \colhead{DEC(1950.0)} &
\colhead{$t_{intg}$\tablenotemark{\dagger}} & \colhead{$rms$\tablenotemark{\dagger}} & \colhead{QF\tablenotemark{\sharp}} & \colhead{survey}\\
\colhead{ } & \colhead{ } & \colhead{($\degr$)} & \colhead{($\degr$)} &
\colhead{(hr min s)} & \colhead{($\degr$ $\arcmin$ $\arcsec$)} &
\colhead{$(hr)$} & \colhead{(mK)} & \colhead{ } & \colhead{ }
}
\startdata
G0.605$+$0.325                      & \nodata          &            0.605 &         $+$0.329 &               17              42            37.7 &            $-$28              13            52.0 &             38.0 &             0.47 &          {\it~E} &           $^3$He \\
G0.665$-$0.035                      & Sgr~B2           &            0.665 &         $-$0.035 &               17              44            10.7 &            $-$28              22            17.0 &              8.4 &             3.28 &          {\it~C} &           $^3$He \\
G1.13$-$0.1                         & \nodata          &            1.119 &         $-$0.098 &               17              45            30.0 &            $-$28              01            01.0 &             55.4 &             1.16 &          {\it~B} &           $^3$He \\
G2.90$+$0.0                         & \nodata          &            2.901 &         $-$0.003 &               17              49            16.7 &            $-$26              26            29.0 &             59.0 &             0.73 &          {\it~D} &           $^3$He \\
G3.270$-$0.101                      & \nodata          &            3.270 &         $-$0.100 &               17              50            29.2 &            $-$26              10            24.0 &             59.9 &             0.83 &          {\it~C} &           $^3$He \\
G5.899$-$0.427                      & \nodata          &            5.900 &         $-$0.427 &               17              57            36.5 &            $-$24              04            16.0 &             16.8 &             3.24 &          {\it~C} &             \cii \\
G5.956$-$1.265                      & \nodata          &            5.956 &         $-$1.264 &               18              00            55.3 &            $-$24              26            21.0 &             12.8 &             3.17 &          {\it~B} &             \cii \\
\enddata
\tablenotetext{\dagger}{$t_{intg}$ and $rms$ correspond to the spectral line data.} 
\tablenotetext{\sharp}{Spectral line quality factor. See text. QF is an evaluation 
of the quality of the parameters measured for the spectra as a whole. It is based on the 
signal-to-noise ratio, the structure of the baseline, the crowding of spectral lines, 
and the accuracy of the Gaussian fit to the line shape.}
\tablenotetext{a}{M16, M17, NGC~6334, Rosette and S209 were observed 
at different positions in the $^3$He survey.}
\tablenotetext{b}{Observed in both \cii\ and $^3$He surveys.}
\label{tab:properties}
\end{deluxetable}

\clearpage
\begin{deluxetable}{lrrrrrrr}
\tabletypesize{\scriptsize}
\tablecaption{Hydrogen Radio Recombination Line Parameters}
\tablewidth{0pt}
\tablehead{
\colhead{Source} & \colhead{$T_L$} & \colhead{$\sigma T_L$} & \colhead{$\Delta v$} & \colhead{$\sigma \Delta v$} &
\colhead{$V$\tablenotemark{\ddagger}} & \colhead{$\sigma V$\tablenotemark{\ddagger}} & \colhead{QF\tablenotemark{\sharp}}\\
\colhead{ } & \colhead{(mK)} & \colhead{(mK)} &  \colhead{(\kms)} & \colhead{(\kms)} &
\colhead{(\kms)} & \colhead{(\kms)} & \colhead{ }
}
\startdata
G0.605$+$0.325                      &            14.92 &             0.58 &            26.03 &             1.17 &          $-$2.88 &             0.50 &          {\it~D} \\
G0.665$-$0.035                      &           474.49 &             4.03 &            40.54 &             0.40 &            64.31 &             0.17 &          {\it~C} \\
G1.13$-$0.1                         &           141.84 &             0.81 &            24.53 &             0.16 &         $-$19.91 &             0.07 &          {\it~B} \\
G2.90$+$0.0                         &            26.76 &             0.38 &            22.98 &             0.37 &          $-$2.08 &             0.16 &          {\it~C} \\
G3.270$-$0.101                      &            50.16 &             0.63 &            25.34 &             0.37 &             7.10 &             0.16 &          {\it~B} \\
G5.899$-$0.427                      &           219.48 &             1.50 &            22.50 &             0.18 &             9.90 &             0.06 &          {\it~B} \\
G5.956$-$1.265                      &           464.75 &             3.90 &            26.15 &             0.25 &             2.61 &             0.07 &          {\it~B} \\
G5.973$-$1.178                      &           526.18 &             4.15 &            26.53 &             0.24 &             2.69 &             0.07 &          {\it~B} \\
\enddata
\tablenotetext{\ddagger}{Note that $V = V_{\rm LSR}$.}
\tablenotetext{\sharp}{Hydrogen line quality factor. See text. QF is a qualitative 
evaluation of the quality of the parameters measured for the H line. Most based 
in the accuracy of the Gaussian fit to the line shape.}
\tablenotetext{a}{M16, M17, NGC~6334, Rosette and S209 were observed 
at different positions in the $^3$He survey.}
\tablenotetext{b}{Observed in the \cii\ survey.}
\tablenotetext{c}{Observed in the $^3$He survey.}
\label{tab:H}
\end{deluxetable}

\clearpage
\begin{deluxetable}{lrrrrrrrrr}
\tabletypesize{\scriptsize}
\tablecaption{Helium Radio Recombination Line Parameters}
\tablewidth{0pt}
\tablehead{
\colhead{Source} & \colhead{$T_L$} & \colhead{$\sigma T_L$} & \colhead{$\Delta v$} & \colhead{$\sigma \Delta v$} &
\colhead{$V$} & \colhead{$\sigma V$} & \colhead{QF\tablenotemark{\sharp}}\\
\colhead{ } & \colhead{(mK)} & \colhead{(mK)} &  \colhead{(\kms)} & \colhead{(\kms)} &
\colhead{(\kms)} & \colhead{(\kms)} & \colhead{ }
}
\startdata
G0.605$+$0.325                      &             5.23 &             1.30 &             4.82 &             1.34 &          $-$3.68 &             0.63 &          {\it~E} \\
G0.665$-$0.035                      &            26.89 &             5.10 &            27.51 &             8.55 &            63.74 &             3.30 &          {\it~B} \\
G1.13$-$0.1                         &            11.26 &             0.91 &            20.22 &             2.03 &         $-$19.96 &             0.81 &          {\it~C} \\
G2.90$+$0.0                         &             1.87 &             0.46 &            17.13 &             1.96 &          $-$8.55 &             2.62 &          {\it~D} \\
G3.270$-$0.101                      &             5.20 &             0.44 &            18.88 &             2.16 &             4.88 &             1.95 &          {\it~D} \\
G5.899$-$0.427                      &            20.75 &             1.35 &            18.85 &             2.00 &             9.17 &             0.61 &          {\it~C} \\
G5.956$-$1.265                      &            46.90 &             1.24 &            19.15 &             0.67 &             0.89 &             0.26 &          {\it~B} \\
G5.973$-$1.178                      &            51.70 &             1.82 &            19.76 &             0.97 &             0.78 &             0.29 &          {\it~C} \\
\enddata
\tablenotetext{\sharp}{Helium line quality factor. See text. QF is a qualitative 
evaluation of the quality of the parameters measured for the He line. Most based 
in the accuracy of the Gaussian fit to the line shape.}
\tablenotetext{a}{M16, M17, NGC~6334, Rosette and S209 were observed 
at different positions in the $^3$He survey.}
\tablenotetext{b}{Observed in the \cii\ survey.}
\tablenotetext{c}{Observed in the $^3$He survey.}
\label{tab:He}
\end{deluxetable}

\clearpage
\begin{deluxetable}{lrrrrrrrrr}
\tabletypesize{\scriptsize}
\tablecaption{Carbon Radio Recombination Line Parameters}
\tablewidth{0pt}
\tablehead{
\colhead{Source} & \colhead{$T_L$} & \colhead{$\sigma T_L$} & \colhead{$\Delta v$} & \colhead{$\sigma \Delta v$} &
\colhead{$V$} & \colhead{$\sigma V$} & \colhead{morphology\tablenotemark{\ast}} & \colhead{RF\tablenotemark{\ast\ast}} & \colhead{QF\tablenotemark{\sharp}}\\
\colhead{ } & \colhead{(mK)} & \colhead{(mK)} & \colhead{(\kms)} & \colhead{(\kms)} &
\colhead{(\kms)} & \colhead{(\kms)} & \colhead{ } & \colhead{ } & \colhead{ }
}
\startdata
G0.605$+$0.325                      &             1.15 &             0.76 &            16.27 &            13.92 &          $-$4.34 &             5.41 &         r &          P &          {\it~E} \\
G0.665$-$0.035                      &             7.00 &             8.88 &            10.88 &            17.42 &            67.19 &             6.89 &         b &          P &          {\it~C} \\
G1.13$-$0.1                         &             1.92 &             1.54 &             6.92 &             6.66 &         $-$15.33 &             2.78 &        sb &          P &          {\it~C} \\
G2.90$+$0.0                         &             2.15 &             0.85 &             6.10 &             2.80 &             0.20 &             1.18 &        sb &          G &          {\it~C} \\
G3.270$-$0.101                      &             3.22 &             0.30 &            10.08 &             1.41 &             1.05 &             0.48 &         r &          F &          {\it~C} \\
G5.899$-$0.427                      &            12.22 &             1.11 &            11.57 &             1.50 &             7.63 &             0.50 &        sb &          G &          {\it~C} \\
G5.956$-$1.265                      &            17.91 &             2.85 &             4.46 &             1.22 &            10.03 &             0.46 &         b &          F &          {\it~C} \\
G5.973$-$1.178                      &            18.89 &             3.60 &             6.05 &             1.45 &             5.99 &             0.48 &         b &          G &          {\it~C} \\
\enddata
\tablenotetext{\ast}{Carbon spectral lineshape morphology: resolved (r), 
semi-blended (sb), or blended (b). See text. It indicates the position 
(velocity) of the C line in the spectra and how distinct it is from the He line.}
\tablenotetext{\ast\ast}{Carbon line reliability factor: excellent (E), good (G), 
fair (F), or poor (P). See text. RF is a qualitative evaluation of the 
reality of a C line.}
\tablenotetext{\sharp}{Carbon line quality factor. See text. QF is a qualitative 
evaluation of the quality of the parameters measured for the C line. Most based 
in the accuracy of the Gaussian fit to the line shape.}
\tablenotetext{a}{M16, M17, NGC~6334, Rosette and S209 were observed 
at different positions in the $^3$He survey.}
\tablenotetext{b}{Observed in the \cii\ survey.}
\tablenotetext{c}{Observed in the $^3$He survey.}
\tablenotetext{d}{Two possible C91$\alpha$ line detections.}
\label{tab:cii}
\end{deluxetable}

\clearpage
\begin{deluxetable}{lrrrrrrrrrrr}
\tabletypesize{\scriptsize}
\rotate
\tablecaption{Radio Emission Continuum Parameters (Total Power Observing Mode)}
\tablewidth{0pt}
\tablehead{
\colhead{Source} & \colhead{$T_C({\rm RA})$} & \colhead{$\sigma T_C({\rm RA})$} &
\colhead{$T_C({\rm DEC})$} & \colhead{$\sigma T_C({\rm DEC})$} &
\colhead{$\Theta({\rm RA})$} & \colhead{$\sigma \Theta({\rm RA})$} &
\colhead{$\Theta({\rm DEC})$} & \colhead{$\sigma \Theta({\rm DEC})$} &
\colhead{$\Theta_{src}$} & \colhead{$\sigma \Theta_{src}$} &
\colhead{QF\tablenotemark{\sharp}}\\
\colhead{ } & \colhead{(K)} & \colhead{(K)} &
\colhead{(K)} & \colhead{(K)} &
\colhead{(arcmin)} & \colhead{(arcmin)} &
\colhead{(arcmin)} & \colhead{(arcmin)} &
\colhead{(arcmin)} & \colhead{(arcmin)} &
\colhead{ }
}
\startdata
G0.605$+$0.325                      &            0.147 &            0.006 &            0.122 &            0.000 &            4.780 &            0.332 &            4.620 &            0.120 &            4.700 &            0.177 &          {\it~B} \\
G0.665$-$0.035                      &            9.039 &            0.008 &            8.469 &            0.033 &            3.686 &            0.006 &            4.858 &            0.046 &            4.232 &            0.021 &          {\it~B} \\
G1.13$-$0.1                         &          \nodata &          \nodata &          \nodata &          \nodata &          \nodata &          \nodata &          \nodata &          \nodata &          \nodata &          \nodata &          \nodata \\
G2.90$+$0.0                         &            0.138 &            0.005 &            0.160 &            0.008 &            5.243 &            0.422 &            6.248 &            0.569 &            5.745 &            0.354 &          {\it~C} \\
G3.270$-$0.101                      &            0.499 &            0.047 &            0.527 &            0.023 &            4.943 &            0.384 &            6.035 &            0.154 &            5.489 &            0.207 &          {\it~B} \\
G5.899$-$0.427                      &          \nodata &          \nodata &          \nodata &          \nodata &          \nodata &          \nodata &          \nodata &          \nodata &          \nodata &          \nodata &          \nodata \\
G5.956$-$1.265                      &            1.990 &            0.003 &            2.006 &            0.004 &            9.229 &            0.251 &            7.411 &            0.132 &            8.270 &            0.135 &          {\it~C} \\
G5.973$-$1.178                      &            5.967 &            0.007 &            5.901 &            0.005 &            4.525 &            0.015 &            4.765 &            0.015 &            4.643 &            0.011 &          {\it~B} \\
\enddata
\tablenotetext{\sharp}{Continuum quality factor. See text. QF is an evaluation 
of the quality of the parameters measured for the continuum scans. It is based on the 
accuracy of the Gaussian fit to the emission feature, the uncertainty in the baseline 
models and the contrast between the target source emission and any emission from
possible sub-structures or nearby confusing sources.}
\tablenotetext{a}{M16, M17, NGC~6334, Rosette and S209 were observed 
at different positions in the $^3$He survey.}
\tablenotetext{b}{Observed in the \cii\ survey.}
\tablenotetext{c}{Observed in the $^3$He survey.}
\label{tab:conttp}
\end{deluxetable}

\clearpage
\begin{deluxetable}{lrrrrrrrrrrr}
\tabletypesize{\scriptsize}
\rotate
\tablecaption{Radio Emission Continuum Parameters (Switched Power Observing Mode)}
\tablewidth{0pt}
\tablehead{
\colhead{Source} & \colhead{$T_C({\rm RA})$} & \colhead{$\sigma T_C({\rm RA})$} &
\colhead{$T_C({\rm DEC})$} & \colhead{$\sigma T_C({\rm DEC})$} &
\colhead{$\Theta({\rm RA})$} & \colhead{$\sigma \Theta({\rm RA})$} &
\colhead{$\Theta({\rm DEC})$} & \colhead{$\sigma \Theta({\rm DEC})$} &
\colhead{$\Theta_{src}$} & \colhead{$\sigma \Theta_{src}$} &
\colhead{QF\tablenotemark{\sharp}} \\
\colhead{ } & \colhead{(K)} & \colhead{(K)} &
\colhead{(K)} & \colhead{(K)} &
\colhead{(arcmin)} & \colhead{(arcmin)} &
\colhead{(arcmin)} & \colhead{(arcmin)} &
\colhead{(arcmin)} & \colhead{(arcmin)} &
\colhead{ }
}
\startdata
G0.605$+$0.325                      &          \nodata &          \nodata &          \nodata &          \nodata &          \nodata &          \nodata &          \nodata &          \nodata &          \nodata &          \nodata &          \nodata \\
G0.665$-$0.035                      &            8.590 &            0.368 &            8.088 &            0.300 &            4.365 &            0.157 &            5.871 &            0.102 &            5.062 &            0.101 &          {\it~A} \\
G1.13$-$0.1                         &            1.311 &            0.016 &            1.340 &            0.014 &            4.837 &            0.091 &            6.415 &            0.088 &            5.570 &            0.065 &          {\it~C} \\
G2.90$+$0.0                         &          \nodata &          \nodata &          \nodata &          \nodata &          \nodata &          \nodata &          \nodata &          \nodata &          \nodata &          \nodata &          \nodata \\
G3.270$-$0.101                      &          \nodata &          \nodata &          \nodata &          \nodata &          \nodata &          \nodata &          \nodata &          \nodata &          \nodata &          \nodata &          \nodata \\
G5.899$-$0.427                      &            3.271 &            0.006 &            3.077 &            0.013 &            5.186 &            0.011 &            4.450 &            0.023 &            4.804 &            0.014 &          {\it~B} \\
G5.956$-$1.265                      &          \nodata &          \nodata &          \nodata &          \nodata &          \nodata &          \nodata &          \nodata &          \nodata &          \nodata &          \nodata &          \nodata \\
G5.973$-$1.178                      &            6.327 &            0.017 &            6.197 &            0.022 &            4.853 &            0.015 &            5.069 &            0.023 &            4.959 &            0.014 &          {\it~B} \\
\enddata
\tablenotetext{\sharp}{Continuum quality factor. See text. QF is an evaluation 
of the quality of the parameters measured for the continuum scans. It is based on the 
accuracy of the Gaussian fit to the emission feature, the uncertainty in the baseline 
models and the contrast between the target source emission and any emission from
possible sub-structures or nearby confusing sources.}
\tablenotetext{a}{M16, M17, NGC~6334, Rosette and S209 were observed 
at different positions in the $^3$He survey.}
\tablenotetext{b}{Observed in the \cii\ survey.}
\tablenotetext{c}{Observed in the $^3$He survey.}
\label{tab:contsp}
\end{deluxetable}

\clearpage

\end{document}